\documentclass[twocolumn,showpacs,showkeys,preprintnumbers,amsmath,amssymb,superscriptaddress]{revtex4-2}
\usepackage{hhline}
\usepackage{bm}
\usepackage{amsmath}
\usepackage{float}
\usepackage{rotating}
\usepackage{hyperref}
\usepackage{dcolumn}
\usepackage{array}
\usepackage{longtable}
\usepackage{graphicx}
\usepackage{booktabs}
\usepackage{booktabs}
\usepackage{multirow}
\usepackage{bm}
\UseRawInputEncoding

\begin{document}

\title{Effects of the strong Breit interaction on the $2s2p - 1s2s$ transitions of inner shell hole states of Helium-like ions}


\author{Xiaobin Ding}
\email[Corresponding author:]{dingxb@nwnu.edu.cn}
\affiliation{Key Laboratory of Atomic and Molecular Physics and Functional Materials of Gansu Province, College of Physics and Electronic Engineering, Northwest Normal University, Lanzhou, 730070, China}
\affiliation{Gansu International Scientific and Technological Cooperation Base of Laser plasma Spectroscopy, Lanzhou 730070, China}

\author{Runxia Zhao}
\affiliation{Key Laboratory of Atomic and Molecular Physics and Functional Materials of Gansu Province, College of Physics and Electronic Engineering, Northwest Normal University, Lanzhou, 730070, China}

\author{Cunqiang Wu}
\affiliation{Key Laboratory of Atomic and Molecular Physics and Functional Materials of Gansu Province, College of Physics and Electronic Engineering, Northwest Normal University, Lanzhou, 730070, China}

\author{Denghong Zhang}
\affiliation{Key Laboratory of Atomic and Molecular Physics and Functional Materials of Gansu Province, College of Physics and Electronic Engineering, Northwest Normal University, Lanzhou, 730070, China}
\affiliation{Gansu International Scientific and Technological Cooperation Base of Laser plasma Spectroscopy, Lanzhou 730070, China}

\author{Mingwu Zhang}
\affiliation{Institute of Modern Physics, Chinese Academy of Sciences, Lanzhou 730000, China}

\author{Yingli Xue}
\affiliation{Institute of Modern Physics, Chinese Academy of Sciences, Lanzhou 730000, China}

\author{Deyang Yu}
\affiliation{Institute of Modern Physics, Chinese Academy of Sciences, Lanzhou 730000, China}

\author{Chenzhonng Dong}
\affiliation{Key Laboratory of Atomic and Molecular Physics and Functional Materials of Gansu Province, College of Physics and Electronic Engineering, Northwest Normal University, Lanzhou, 730070, China}
\affiliation{Gansu International Scientific and Technological Cooperation Base of Laser plasma Spectroscopy, Lanzhou 730070, China}

\thanks{}


\date{\today}

\begin{abstract}
We have calculated the transition energies and probabilities of one-electron one photon and one-electron two photon transitions of middle-Z and high-Z He-like ions using the fully relativistic multiconfiguration Dirac-Hartree-Fock method with active space method. The relativistic, electron correlation, Breit and QED effects are systemically taken into account in the present work. Results showcase consistent agreement with the experimental and theoretical data, uncovering intriguing inversion phenomena in One-Electron One-Photon transitions energy, particularly in double-hole states. Theoretical spectra intensities provide valuable insights into high-energy X-ray radiation processes from double \textit{K}-hole states. 
\end{abstract}


\maketitle

\section{INTRODUCTION}

The Breit interactions between electrons in atoms or ions have been of significant interest for decades, dating back to Gregory Breit's proposal of the two-electron interactions theory for studying the fine-structured energy levels of helium atoms \cite{Breit1929,Breit1930}. The Breit interaction, representing the lowest-order correction to the Coulomb interaction between two electrons, involves exchanging a transverse photon \cite{Dyall1989}.

Extensive research in the field of atomic physics has explored the effects of Breit interactions on fine- and hyperfine-structure energy levels \cite{Si2018,Dinh2016,Malyshev2014,Safronova2014,Aourir2018,Lochmann2014,Sushkov2001,Johnson1997}, radiative and non-radiative decays \cite{Chen1987,Bilal2018,Kasthurirangan2013,Korobov2003,Ding2021}, and multiplet splittings of double-inner-shell-vacancy configurations in heavy atoms \cite{Mann1971}. Notably, Joseph B. Mann et al. found that for high-Z atomic ions, the contribution of Breit interaction to total energy remains relatively small \cite{Mann1971}. However, Mau Hsiung Chen et al. discovered its significant effect on multiplet splittings in heavy atoms \cite{Chen1982}.

Subsequent studies revealed the varying impact of Breit interactions on different atomic properties. For instance, L. Natarajan et al. found that Breit interactions either enhance or reduce the transition rates for Li-like ions \cite{Natarajan2010,Natarajan2015}. E. A. Konovalova et al. analyzed the influence of Breit interaction on the energy levels of Mg isoelectronic series, noting its importance for heavy ions \cite{Konovalova2015}. Additionally, Ding et al. highlighted the dominance of electron correlation for low Z elements, while Breit interaction becomes more significant for high-Z ions \cite{Ding2020}.

In the contemporary landscape, many theoretical calculations of atomic structures and transition properties incorporate Breit interactions. The energy level structure and transition properties of inner shell hollow atoms have become a central focus \cite{Hoogkamer1976,Briand1976,Nagel1976,Aaberg1976,Stoller1976,Safronova1977}. Inner shell hollow atoms, characterized by the occupation of outer shell orbitals while the inner shell remains unoccupied, undergo deexcitation primarily through non-radiative Auger processes and high-energy X-ray photon radiation processes \cite{Inhester2012}. Despite the inherent instability of these exotic atoms, the study of high-energy X-ray radiation processes offers a convenient and highly precise analytical approach \cite{Porquet2010,Widmann2000}.

Particularly noteworthy is the radiative deexcitation of double \textit{K} shell hole states, where the two-electron one-photon (TEOP) transition plays a pivotal role. This transition, strictly forbidden in an independent electron approximation, was first predicted by Heisenberg in 1925 \cite{Heisenberg1985}, with subsequent identification of selection rules by Goudsmit in 1931 \cite{Goudsmit1931}. Experimental validation came from Wolfi et al., who observed TEOP transition spectra for the first time in Ni-Ni, Ni-Fe, Fe-Ni, and Fe-Fe collision experiments \cite{Woelfli1975}.

The current research aims to contribute to understanding the influence of electron correlation effects, Breit interaction, and quantum electrodynamics (QED) effects on the energy level structure and radiative transition properties of inner shell hole state atoms. Notable experimental efforts in recent decades include the observation of two-electron one-photon \textit{K}-shell double ionization \cite{Hoszowska2011} and the resonant excitation of the \textit{K}-series of He-like and Li-like oxygen ions, revealing a strong two-electron one-photon transition \cite{Togawa2020}. Konstantin N. Lyashchenko's evaluation of TEOP transition properties within the QED theory further advanced our understanding \cite{Lyashchenko2021}.

However, there remains a gap in understanding the effect of Breit interaction on the decay process of inner shell hole states, either by one-electron one-photon (OEOP) or two-electron one-photon (TEOP) transition. The present work focuses on studying the transition energy and probabilities for TEOP and OEOP for various He-like ions, including N$^{5+}$, Ne$^{8+}$, Fe$^{24+}$, Cu$^{27+}$, Zn$^{28+}$, Ga$^{29+}$, Ge$^{30+}$, As$^{31+}$, Se$^{32+}$, Br$^{33+}$, Kr$^{34+}$, Xe$^{52+}$ and W$^{72+}$, using multi-configuration Dirac-Hartree-Fock (MCDF) method. Our calculations include an analysis of the effect of Breit interaction on the transition energy, considering the relativistic, electron correlation effects, and the finite nuclear size effect with a two-parameter Fermi charge distribution. The results are obtained using the GRASP2K \cite{Fischer2019} code and are expected to provide valuable insights for future experimental and theoretical endeavors.

\section{Theory}

The energy level structure and transition properties are calculated using the MCDF method, including the Breit interaction, self-energy, and vacuum polarization, along with the active space method \cite{Ding2014,Ding2011,Costa2006,Martins2004}. The MCDF method has demonstrated efficiency in addressing electron correlation effects arising from the strong interaction of nearly degenerate excited states with the reference state \cite{Natarajan2009,Natarajan2007}. Detailed descriptions of the method can be found in Grant's monograph \cite{Grant2007}. Only a brief description is given below.

A linear combination of configuration state functions (CSFs) is used in the construction of atomic state functions (ASFs) with the same parity $P$, total angular momentum $J$, and total magnetic quantum number $M$:

\begin{equation}
	\Psi(\mathrm{PJM}_J)=\sum_{i=1}^{N_c} C_i\phi(\gamma_i\mathrm{PJM}_J)
\end{equation}
where $N_c$ is the number of CSFs, $C_i$ is the mixing coefficient for state $i$, and $\gamma_i$ encompasses all the one-electron and intermediate quantum numbers necessary to define the CSFs. In a multi-configuration relativistic calculation, CSFs are symmetry-adapted linear combinations of Slater determinants constructed from a set of one-electron Dirac spinors \cite{Fischer2016}. The radial wave function and mixing coefficient are determined using the extended level optimization model in self-consistent field calculations. Specifically, diagonalize the Dirac-Coulomb Hamiltonian as follows:

\begin{equation}
	H_{DC}=\sum_i[c\bm{\alpha}_i \cdot \bm{p}_i + (\beta-1)c^2 - \frac{Z}{r}] + \sum_{i>j} \frac{1}{r_{ij}}
\end{equation}
where $-\frac{Z}{r}$ is the electron-nucleus Coulomb interaction, $\bm{\alpha}$ and $\beta$ the 4$\times$4 Dirac matrices, and $c$ the speed of light in atomic unit. 

Relativistic configuration interaction calculations can be accomplished by incorporating higher-order interactions, such as Breit interaction, self-energy, and vacuum polarization etc., in the Hamiltonian.

It is crucial for us to focus on the correction of electron-electron interaction. This term is a sum of the Coulomb interaction $\hat{V}^{C}_{ij}$ operator and the transverse Breit $\hat{V}^{B}_{ij}$ operator:

\begin{equation}
	\hat{V}_{ij}=\hat{V}^{C}_{ij}+\hat{V}^{B}_{ij}
\end{equation}
where the Coulomb interaction operator is $\hat{V}^{C}_{ij}=\frac{1}{r_{ij}}$, and the Dirac-Colulom-Breit operator in the Coulomb gauge is:

\begin{equation}
	\begin{split}
	\hat{V}^{B}_{ij}=-\bm{\alpha}_i \cdot \bm{\alpha}_j & \frac{\text{cos}(\omega_{ij}r_{ij}/c)}{r_{ij}}  \\
	&-  (\bm{\alpha}_i \cdot \bm{\nabla}_i)(\bm{\alpha}_j \cdot \bm{\nabla}_j) \frac{\text{cos}(\omega_{ij} r_{ij} /c)-1}{\omega^{2}_{ij}{r_{ij}/c^2}}
	\end{split}
\end{equation}
where $\omega_{ij}$ is the frequency of the virtual photon exchanged. The unretarded (instantaneous) parts are obtained by applying the long wavelength (zero-frequency) approximation $\omega_{ij}\rightarrow0$. Then, the Breit interaction terms in the Coulomb gauge are given as:

\begin{equation}\label{eq5}
	\begin{aligned}
		\hat{V}^{B}_{ij}&=-\frac{\bm{\alpha}_i \cdot \bm{\alpha}_j}{r_{ij}}+(\frac{\bm{\alpha}_i \cdot \bm{\alpha}_j}{2r_{ij}}-\frac{(\bm{\alpha}_i \cdot r_{ij})(\bm{\alpha}_j \cdot r_{ij})}{2r^{3}_{ij}})
	\end{aligned}
\end{equation}
where the first term is called the Gaunt (magnetic) part, and the second part is known as the retardation part. The zero-frequency approximation to the full transverse Breit interaction, i.e., Eq.(\ref{eq5}), is well-suited for most computations of many-electron atomic systems \cite{Gorceix1987,Grant1976,Lindroth1989}.

Moreover, we utilize the active space method to consider the electron correlation effects in the multi-electron system. In the first step, the initial orbit is obtained by single-configuration Dirac--Hartree-Fock calculation. Next, we systematically extend the active space by allowing single and double (SD) electron excitation from the reference sets. We also assess the eigenenergy and transition parameters with the optimized wave function. Then, the size of the active space is gradually enlarged in the principal quantum number $n$ layer by layer until observable convergence and stability are obtained. The active set considered in this work consisted of relativistic subshells with the principal quantum number $n\leq7$. The active space is extended to the first layer, namely $n=3$, $l\le2$ ($n3l2$) virtual orbit, where $l$ is the angular moment quantum number. All newly added correlation orbits are optimized, keeping the previously optimized orbits unchanged. Our calculation found that the contribution of correlation orbitals from higher $n$ and $l$ is tiny, so we limit the active space to $n7l4$.

\section{RESULT AND DISCUSSION}


%



\setlength{\tabcolsep}{4.0mm}{
	\begin{table*}
		\small
		\caption{The eigenenergy and excitation energy (in eV) for the excited states $1s2s$ and $2s2p$ configurations of N$^{5+}$, Fe$^{24+}$ and Xe$^{52+}$ ions. DF denotes the initial Dirac-Fock calculation and  $nalb$ represent the active sets consisting of all orbitals from $n \le a$ and $l \le b$. E$_R$ (in eV) is the excitation energy relative to the ground state $1s^2$ with the relativistic and electron correlation effects, and E$_{RCI}$ (in eV) is the excitation energy additionally including the Breit interaction and QED effect.} \label{Tab1}
		\centering
		\begin{tabular}{cccccccccccc}
			\hline
			&&&\multicolumn{3}{c}{He-like N}\\
			\hline
			\multicolumn{2}{c}{\multirow{2}{*}{Active sets}} & \multicolumn{4}{c}{2s2p} && \multicolumn{2}{c}{1s2s} \\
			\cline{3-6}
			\cline{8-9}\noalign{\smallskip}
			\multicolumn{2}{c}{} & $^{3}P_{0}$ & $^{3}P_{1}$ & $^{3}P_{2}$ & $^{1}P_{1}$ & & $^{3}S_{1}$ & $^{1}S_{0}$ \\
			\hline
			\multicolumn{2}{c}{ DF }   &-308.80 &-308.77 & -308.71 & -297.21  &             &-799.30      &-792.85\\
			\multicolumn{2}{c}{ $n3l2$ }&-308.77 &-308.74 &  -308.67 & -298.27&             &-799.22      &-791.98\\
			\multicolumn{2}{c}{ $n4l3$ } &-308.81 &-308.78 &  -308.71 &-298.46&             &-799.23      &-792.07\\
			\multicolumn{2}{c}{ $n5l4$ } &-308.89 &-308.86 &  -308.80 &-299.04&             &-799.24      &-792.11\\
			\multicolumn{2}{c}{ $n6l4$ } &-308.90 &-308.87 &  -308.81 &-299.05&             &-799.24      &-792.13\\
			\multicolumn{2}{c}{ $n7l4$ } &-308.90 &-308.87 &  -308.81 &-299.06&             &-799.24      &-792.14\\
			\multicolumn{2}{c}{E$_R$}    &909.89  &909.92  &  909.98  &919.73 &             &419.55       &426.66 \\
			\multicolumn{2}{c}{E$_{RCI}$} &909.68 &909.70 &  909.76   &919.51 &             &419.40       &426.53\\ 
			\multicolumn{2}{c}{Ref\cite{Kadrekar2011}}&909.49  &909.52  &  909.59  &919.17 & &419.17      &425.91\\
			\multicolumn{2}{c}{NIST\cite{Kramida2023}}&       &        &        &        &   &419.80      &426.42 \\
			\hline
			&&&\multicolumn{3}{c}{He-like Fe}\\
			\hline
			\multicolumn{2}{c}{\multirow{2}{*}{Active sets}} & \multicolumn{4}{c}{2s2p} && \multicolumn{2}{c}{1s2s} \\
			\cline{3-6}
			\cline{8-9}\noalign{\smallskip}
			\multicolumn{2}{c}{} & $^{3}P_{0}$ & $^{3}P_{1}$ & $^{3}P_{2}$ & $^{1}P_{1}$ & & $^{3}S_{1}$ & $^{1}S_{0}$ \\
			\hline
			\multicolumn{2}{c}{ DF }  &-4556.48 &-4551.52 &-4535.97 &-4499.37 &              &-11473.68    &-11443.80\\
			\multicolumn{2}{c}{$n3l2$} &-4556.44 &-4551.50 &-4535.93 &-4499.66&              &-11473.61    &-11442.57\\
			\multicolumn{2}{c}{$n4l3$} &-4556.49 &-4551.55 &-4535.98 &-4499.89&              &-11473.62    &-11442.66\\
			\multicolumn{2}{c}{$n5l4$} &-4556.57 &-4551.65 &-4536.08 &-4500.52&              &-11473.62    &-11442.69\\
			\multicolumn{2}{c}{$n6l4$} &-4556.58 &-4551.66 &-4536.08 &-4500.54&              &-11473.62    &-11442.71\\
			\multicolumn{2}{c}{$n7l4$} &-4556.58 &-4551.66 &-4536.09 &-4500.55&              &-11473.62    &-11442.72\\
			\multicolumn{2}{c}{E$_R$} &13562.49 &13567.41 &13582.98  &13618.52&              &6645.45      &6676.35\\
			\multicolumn{2}{c}{E$_{RCI}$}&13548.78 &13553.55 &13569.19 &13604.78&            &6635.81      &6667.95\\
			\multicolumn{2}{c}{Ref\cite{Kadrekar2011}} & 13549.05 & 13553.81 & 13569.47  &13604.88& &6635.71 &6667.39\\
			\multicolumn{2}{c}{NIST\cite{Kramida2023}}&13550.4  &13555.1  &13570.4&   13605.7&      &6636.60  &6668.02\\
			\hline
			&&&\multicolumn{3}{c}{He-like Xe}\\
			\hline
			\multicolumn{2}{c}{\multirow{2}{*}{Active sets}} & \multicolumn{4}{c}{2s2p} && \multicolumn{2}{c}{1s2s} \\
			\cline{3-6}
			\cline{8-9}\noalign{\smallskip}
			\multicolumn{2}{c}{} & $^{3}P_{0}$ & $^{3}P_{1}$ & $^{3}P_{2}$ & $^{1}P_{1}$ & & $^{3}S_{1}$ & $^{1}S_{0}$ \\
			\hline
			\multicolumn{2}{c}{DF}   &-20676.89 &-20651.29 &-20256.91 &-20192.60&          &-51492.74     &-51459.78\\
			\multicolumn{2}{c}{$n3l2$} &-20676.82 &-20651.28 &-20256.88&-20192.82&          &-51492.63    &-51418.19\\
			\multicolumn{2}{c}{$n4l3$} &-20676.87 &-20651.38 &-20256.92&-20193.00&          &-51492.64    &-51418.28\\
			\multicolumn{2}{c}{$n5l4$} &-20676.89 &-20651.41 &-20256.94&-20193.07&          &-51492.65    &-51418.32\\
			\multicolumn{2}{c}{$n6l4$} &-20676.90 &-20651.43 &-20256.95&-20193.10&          &-51492.65    &-51418.35\\
			\multicolumn{2}{c}{$n7l4$} &-20676.90 &-20651.45 &-20256.96&-20193.13&          &-51492.65    &-51418.36\\
			\multicolumn{2}{c}{E$_R$} &61038.07  &61063.53  &61458.01 &61521.85&            &30222.32     &30296.61\\
			\multicolumn{2}{c}{E$_{RCI}$} &60898.36 &60922.05 &61317.54&61382.39&           &30124.68     &30211.08\\
			\hline
		\end{tabular}
\end{table*}}

We verified the accuracy of the present calculations by computing and observing the convergence of energy eigenvalues and the excitation energies of $2s2p$ and $1s2s$ states of selected N$^{5+}$, Fe$^{24+}$, and Xe$^{52+}$ He-like ions, and comparing with existing data, displayed in Table \ref{Tab1}. Here, E$_R$ indicates the excitation energy relative to the $1s^2$ ground state and incorporates relativistic and electron correlation effects, while E$_{RCI}$, which includes additional corrections from the Breit interaction and QED effects. The table displays different electron correlation models as active sets and shows a tendency for energy eigenvalues to converge with increased active space. Our calculated excitation energies align well with extant data. For instance, the maximum deviation of the energy eigenvalues of N$^{5+}$ and Fe$^{24+}$ from the results calculated by the MCDF method in the reference \cite{Kadrekar2011} is approximately 0.15\%. Compared to NIST data \cite{Kramida2023}, the maximum deviation is also 0.15\%. Our work's electron correlation model encapsulates the most significant electron correlation effects in He-like systems.

In Table \ref{Tab2}, we present the calculated OEOP transition energies and comparison data for the $2s2p-1s2s$ transitions of He-like ions, including N, Ne, Fe, Cu, Zn, Ga, Ge, As, Se, Br, Kr, Xe, and W. To the author’s knowledge, only N$^{5+}$, Ne$^{8+}$, and Fe$^{24+}$ have experimental comparative data. Therefore, we provide the average relative deviation between our calculated results and experimental observations, which ranges from approximately 0.07\% to 0.16\% \cite{Nicolosi1977, Kroupp2003, Phillips2004, Nandi2008}. Our theoretical transition energies are in excellent agreement with the existing experimental data.

\setlength{\tabcolsep}{4.0mm}{
	\begin{table*}
		\small
		\caption{The transition energy (in eV) for the $2s2p -1s2s$ OEOP transition of the He-like ions. \label{Tab2}}
		\centering
		\begin{tabular}{ccccccccccccccc}
			\hline
			\multicolumn{2}{c}{\multirow{2}{*}{Z}} & \multicolumn{6}{c}{Energy(eV)} \\
			\cline{3-8}\noalign{\smallskip}
			\multicolumn{2}{c}{} & $^{3}P_{1}$ - $^{1}S_{0}$ &  $^{3}P_{0}$ - $^{3}S_{1}$ &$^{3}P_{1}$ - $^{3}S_{1}$ & $^{3}P_{2}$ - $^{3}S_{1}$ &$^{1}P_{1}$ - $^{1}S_{0}$ & $^{1}P_{1}$ - $^{3}S_{1}$  \\
			\hline
			\multicolumn{1}{c}{\multirow{5}{*}{N}}\\
			& Calc.  & 483.20 &490.30 & 490.33 & 490.39  & 493.01 &500.14 \\
			& Ref\cite{Nicolosi1977}  & &  &490.85 &  & 493.80& \\
			& Ref\cite{Kadrekar2011} & 483.61 & 490.32 & 490.35  & 490.42 & 493.26   & 500.00 \\
			& Ref\cite{Goryayev2006} & 483.72 & 490.30 & 490.33  & 490.40 & 493.04 & 499.66\\
			\multicolumn{1}{c}{\multirow{5}{*}{Ne}}\\
			& Calc.  & 996.36 &  1007.04 & 1007.17 & 1007.45 &1011.05 & 1021.87 \\
			& Ref\cite{Kroupp2003}  & & &1007.86  & & & \\
			& Ref\cite{Kadrekar2011} & 996.79 &1007.07 & 1007.20 &  1007.48 & 1011.32 &1021.73  \\
			& Ref\cite{Goryayev2006} & 996.92 &  1007.0 & 1007.2 &1007.5 & 1011.2 & 1021.4 \\
			\multicolumn{1}{c}{\multirow{5}{*}{Fe}}\\
			& Calc.  & 6885.95 & 6913.32 & 6918.09  & 6933.73&6937.19 &  6969.33\\
			& Ref\cite{Phillips2004}  & & &  & &6942   & \\
			& Ref\cite{Kadrekar2011} & 6886.41 & 6913.32 & 6918.10 & 6933.77 & 6937.52 & 6969.20 \\
			& Ref\cite{Goryayev2006} & 6886.7 & 6913.4 & 6918.1& 6933.8 & 6937.6 & 6969.0 \\
			\multicolumn{1}{c}{\multirow{4}{*}{Cu}}\\
			& Calc.  & 8592.77  &8622.80 & 8629.42 & 8654.69& 8656.70 &  8693.35 \\
			& Ref\cite{Shirai1997} & 8594.53 &  8623.87 & 8630.44 & 8654.49 &8657.52 & 8693.42  \\
			& Ref\cite{Lyashchenko2021} & 8593.6  &  &   8629.5 & & 8657.3 & 8693.2 \\
			\multicolumn{1}{c}{\multirow{4}{*}{Zn}}\\
			& Calc.  &9205.03 & 9235.92 & 9243.23 & 9272.57 & 9274.44 & 9312.65 \\
			& Ref\cite{Lyashchenko2021}  & 9205.7 & & 9243.1  &  & 9274.5 & 9311.9 \\
			& Ref\cite{Goryayev2006} & 9205.8 & 9235.9 & 9243.2 & 9272.6& 9274.5 & 9311.9 \\
			\multicolumn{1}{c}{\multirow{4}{*}{Ga}}\\
			& Calc & 9838.98 & 9870.77 & 9878.77 &9912.63&9913.61 &  9953.39 \\
			& Ref\cite{Lyashchenko2021}  & 9839.5  & & 9878.4&  & 9913.9 & 9952.8  \\
			& Ref\cite{Goryayev2006} & 9839.8 &9870.8 & 9878.7 & 9912.7& 9914.0 &  9953.0 \\
			\multicolumn{1}{c}{\multirow{4}{*}{Ge}}\\
			& Calc. &10494.78 &  10527.55 & 10536.21 & 10575.28&10575.44 & 10616.87 \\
			& Ref\cite{Lyashchenko2021} & $10495$ & & $10536$&  &  $10576$ &$10616$ \\
			& Ref\cite{Goryayev2006} & 10496 & 10528 & 10536  & 10575& 10576 & 10617\\
			\multicolumn{1}{c}{\multirow{4}{*}{As}}\\
			& Calc. & 11172.58 & 11206.28 & 11215.64 & 11260.43& 11259.91 & 11302.98 \\
			& Ref\cite{Lyashchenko2021} & 11174 &&11216 &  11261 & & 11303 \\
			& Ref\cite{Goryayev2006} & 11173 & 11206 & 11216 & 11260 &11261 &  11303 \\
			\multicolumn{1}{c}{\multirow{4}{*}{Se}}\\
			& Calc. & 11872.54 &11907.16 & 11917.27 & 11968.38 & 11967.51 &  12012.24\\
			& Ref\cite{Lyashchenko2021} & 11873 &  & 11917& &11968 & 12012 \\
			& Ref\cite{Goryayev2006} & 11873 & 11907 & 11917  & 11968 &11968 &  12011\\
			\multicolumn{1}{c}{\multirow{4}{*}{Br}}\\
			& Calc. & 12594.49 & 12630.13 & 12640.92  &12699.09& 12697.11 & 12743.54\\
			&Ref\cite{Lyashchenko2021} & 12595 & & 12641 & &  12698 & 12743 \\
			& Ref\cite{Goryayev2006} &  12595 & 12630 & 12641  & 12699 & 12698 & 12743 \\
			\multicolumn{1}{c}{\multirow{4}{*}{Kr}}\\
			& Calc. & 13338.82 &13375.47 & 13386.98 & 13452.91& 13450.15 &  13498.31  \\
			& Ref\cite{Lyashchenko2021} & 13339 &  & 13386    && 13450  & 13498  \\
			& Ref\cite{Goryayev2006} & 13340 & 13376 & 13387 & 13453& 13451 & 13498  \\
			\multicolumn{1}{c}{\multirow{4}{*}{Xe}}\\
			& Calc.  & 30712.53 &30775.24 & 30798.94 &31194.45 &31172.90 &  31259.30 \\
			& Ref\cite{Lyashchenko2021} & 30713 & &30798 &   &31173&\\
			\multicolumn{1}{c}{\multirow{4}{*}{W}}\\
			& Calc.  & 59883.02 & 60002.84 & 60036.93 & 61645.39&61583.47 &  61737.39  \\
			& Ref\cite{Lyashchenko2021} & 59881 & & 60031 & & 61581 & 61732 \\
			\hline
		\end{tabular}\\
\end{table*}}

In the current calculation, we also compare with other theoretical data. For N$^{5+}$, Ne$^{8+}$, and Fe$^{24+}$ ions, the difference between our results and those calculated by Kadrekar R et al. with the MCDF method ranges from 0.00014\% to 0.08\% \cite{Kadrekar2011}. The most significant deviation from the results calculated by F.F. Goryayev et al. based on the Z-expansion method is about 0.11\%, specifically for the $^3P_1$-$^1S_0$ transition \cite{Goryayev2006}. The deviation of the remaining results is less than 0.0006\%. For Cu$^{27+}$ ion, the maximum deviation compared with NIST database results \cite{Shirai1997, Kramida2023} is 0.02\%, and the maximum deviation compared with the results calculated by Konstantin N. Lyashchenko et al. using QED theory is about 0.01\% \cite{Lyashchenko2021}. The difference between the calculations of the other ions and those of Konstantin N. Lyashchenko et al. is in the range of 0.005\% to 0.013\%, and that of F. F. Goryayev et al. is in the scope of 0.003\% to 0.012\%. From the above analysis, we can see that our results agree with the results of other authors’ works, providing references for future theories and experiments.

\setlength{\tabcolsep}{4.0mm}{
	\begin{table*}
		\small
		\caption{The transition probability (in $s^{-1}$) for the $2s2p$-$1s2s$ OEOP transition of the He-like ions. $a(b)$ represent $a\times 10^b$\label{Tab3}}
		\centering
		\begin{tabular}{cccccccccccccccc}
			\hline
			\multicolumn{2}{c}{\multirow{2}{*}{Z}} & \multicolumn{6}{c}{$A_{ij} \; (s^{-1})$} \\
			\cline{3-8}\noalign{\smallskip}
			&  & $^{3}P_{1}$ - $^{1}S_{0}$ &  $^{3}P_{0}$ - $^{3}S_{1}$ &$^{3}P_{1}$ - $^{3}S_{1}$ & $^{3}P_{2}$ - $^{3}S_{1}$ &$^{1}P_{1}$ - $^{1}S_{0}$ & $^{1}P_{1}$ - $^{3}S_{1}$  \\
			\hline
			\multicolumn{1}{c}{\multirow{4}{*}{N}}\\
			& Calc. & 2.16(7) &1.31(12) & 1.31(12) & 1.31(12)& 1.29(12) &  1.21(7)   \\
			&Ref\cite{Kadrekar2011} & 3.724(7) &1.330(12) & 1.330(12) & 1.328(7) & 1.401(12) &  2.880(7)    \\
			& Ref\cite{Goryayev2006} & 3.48(7)  & 1.34(12) & 1.34(12) & 1.34(12)&1.42(12) &  3.69(7)   \\
			\multicolumn{1}{c}{\multirow{4}{*}{Ne}}\\
			& Calc. & 1.04(9)   &5.71(12) & 5.71(12) & 5.70(12)& 5.65(12) &8.40(8) \\
			& Ref\cite{Kadrekar2011} & 1.246(9) &  5.755(12) & 5.751(12)   & 5.751(12)&5.964(12) & 1.046(9) \\
			& Ref\cite{Goryayev2006} & 1.17(9) &5.79(12)  & 5.80(12)& 5.744(12)  & 6.02(12) &  1.22(9) \\
			\multicolumn{1}{c}{\multirow{5}{*}{Fe}}\\
			& Calc. & 1.07(13) & 2.78(14) & 2.66(14) & 2.74(13)& 2.63(14) &  1.01(13) \\
			& Ref\cite{Kadrekar2011} & 1.100(13) & 2.780(14) & 2.666(14) & 2.744(14) & 2.683(14) & 1.025(13) \\
			& Ref\cite{Goryayev2006} & 1.06(13) &2.84(14) & 2.74(14)  & 2.87(14) & 2.79(14) &  1.08(13)\\
			\multicolumn{1}{c}{\multirow{5}{*}{Cu}}\\
			& Calc. & 2.66(13) & 4.32(14) & 4.04(14) & 4.24(14)&3.99(14) &  2.52(13)  \\
			& Ref\cite{Goryayev2006} & 2.62(13) & 4.44(14) & 4.18(14)  & 4.49(14)&4.26(14) &  2.69(13)\\
			& Ref\cite{Lyashchenko2021} & 2.73(13) & & 4.05(14) &&  4.06(14) & 2.56(13) \\
			\multicolumn{1}{c}{\multirow{5}{*}{Zn}}\\
			& Calc. & 3.44(13) &  4.95(14) & 4.59(14) & 4.86(14)&4.53(14) & 3.26(13) \\
			& Ref\cite{Goryayev2006} & 3.43(13) & 5.10(14) & 4.77(14)& 5.16(14) & 4.86(14) & 3.52(13)  \\
			& Ref\cite{Lyashchenko2021} & 3.57(13) & &   4.60(14) &&4.60(14) &3.35(13)\\
			\multicolumn{1}{c}{\multirow{5}{*}{Ga}}\\
			& Calc. & 4.45(13) &5.65(14) & 5.19(14) & 5.54(14)& 5.12(14) &  4.22(13)  \\
			& Ref\cite{Goryayev2006} & 4.43(13) &5.83(14) & 5.40(14)  & 5.91(14)& 5.50(14) &  4.55(13)\\
			& Ref\cite{Lyashchenko2021} & 4.60(13) & & 5.20(14) & & 5.20(14) & 4.31(13)\\
			\multicolumn{1}{c}{\multirow{5}{*}{Ge}}\\
			& Calc. & 5.71(13) & 6.42(14) & 5.84(14)& 6.29(14) & 5.76(14) & 5.41(13) \\
			& Ref\cite{Goryayev2006} & 5.64(13)  & 6.64(14) & 6.09(14) & 6.73(14)& 6.21(14)  & 5.79(13)     \\
			& Ref\cite{Lyashchenko2021} & 5.84(13) & & 5.85(14) & & 5.84(14) &5.48(13)  \\
			\multicolumn{1}{c}{\multirow{5}{*}{As}}\\
			& Calc. & 7.17(13) &7.27(14) & 6.54(14) & 7.11(14)& 6.44(14) &  6.79(13) \\
			& Ref\cite{Goryayev2006} & 7.09(13)  &7.53(14) & 6.84(14) & 7.64(14)& 6.98(14) &  7.28(13)  \\
			& Ref\cite{Lyashchenko2021} & 7.32(13) & & 6.55(14) &&   6.53(14) &6.87(13) \\
			\multicolumn{1}{c}{\multirow{5}{*}{Se}}\\
			& Calc. & 8.82(13) &  8.21(14) & 7.31(14) & 8.01(14)&7.19(14) &  8.35(13) \\
			& Ref\cite{Goryayev2006} & 8.81(13)  & 8.52(14) & 7.66(14)& 8.65(14) & 7.81(14) & 9.05(13)  \\
			& Ref\cite{Lyashchenko2021} & 9.06(13) & & 7.31(14) & &7.28(14) & 8.50(13)  \\
			\multicolumn{1}{c}{\multirow{5}{*}{Br}}\\
			& Calc. & 1.09(14) &  9.22(14) & 8.12(14)  & 9.00(14)&7.98(14) & 1.03(14)\\
			& Ref\cite{Goryayev2006} & 1.08(14) &9.60(14) & 8.54(14)  & 9.75(14)& 8.72(14) &  1.11(14) \\
			& Ref\cite{Lyashchenko2021} & 1.11(14) &  & 8.13(14) &  &8.08(14) & 1.04(14) \\
			\multicolumn{1}{c}{\multirow{5}{*}{Kr}}\\
			& Calc. & 1.32(14) &1.03(15) & 9.00(14) & 1.01(15)& 8.83(14) &  1.25(14) \\
			& Ref\cite{Goryayev2006} & 1.31(14) & 1.08(15) & 9.49(14) & 1.10(15)& 9.69(14) & 1.35(14) \\
			& Ref\cite{Lyashchenko2021} & 1.34(14) & & 9.01(14) & &  8.94(14) & 1.26(14)\\
			\multicolumn{1}{c}{\multirow{4}{*}{Xe}}\\
			& Calc. & 1.34(15) &  5.32(15) & 3.98(15) & 5.01(15)&3.78(15) & 1.24(15) \\
			& Ref\cite{Lyashchenko2021} & 1.36(15) & & 3.98(15) &&   3.81(15) &1.25(15) \\
			\multicolumn{1}{c}{\multirow{4}{*}{W}}\\
			& Calc. & 5.78(15) &1.92(16) & 1.34(16) & 1.71(16)&  1.20(16) & 5.09(15) \\
			& Ref\cite{Lyashchenko2021} & 5.81(15) & & 1.34(16) && 1.21(16) & 5.15(15)  \\
			\hline
		\end{tabular}\\
\end{table*}}

The relativistic Babushkin and Coulomb gauges correspond to the length and velocity gauges in nonrelativistic limitations. When using the strict wave function, the transition probabilities of these two gauges should be consistent. Thus, we use the ratio of transition probabilities under the two gauges to judge the quality of the wave function and the accuracy of the calculated data. In most calculations, this ratio is close to 1.0. In Table \ref{Tab3}, we present the transition probability of $2s2p-1s2s$ under the Babushkin gauge and compare them with the results from other works.

As observed in Table \ref{Tab3}, our calculation results align well with the work of Kadrekar R et al. \cite{Kadrekar2011}, where the MCDF and active space methods were employed. Notably, the present probability for Fe compares favorably with their work, and the deviation should not exceed 2.7\%. The discrepancy is attributed to the differences in the orbit set of the extended active space. Our results are further compared with those obtained by Konstantin N. Lyashchenko et al. \cite{Lyashchenko2021}, and the deviation falls within the range of 0-3.26\%. Among them, the results of Br, Kr, Xe, and W exhibit the best agreement. Compared with the Z expansion transition rate \cite{Goryayev2006}, their deviation range is 0.04-17.8\%.

\setlength{\tabcolsep}{4.0mm}{
	\begin{table*}
		\small
		\caption{Transition energy (in eV) and probability (in $s^{-1}$) of $2s2p$ - $1s^2$ TEOP transition of He-like ion. $a(b)$ for the value of $A_{ij}$ represent $a\times 10^b$. \label{Tab4}}
		\centering
		\begin{tabular}{cccccccccccc}
			\hline
			\noalign{\smallskip}
			\multicolumn{2}{c}{\multirow{2}{*}{Z}} & \multicolumn{2}{c}{$^3P_1$ - $^1S_0$} & & \multicolumn{2}{c}{$^1P_1$ -$^1S_0$} \\
			\cline{3-4}
			\cline{6-7}\noalign{\smallskip}
			\multicolumn{2}{c}{} & Energy (eV) & $A_{ij} (s^{-1})$  & & Energy (eV)  & $A_{ij} (s^{-1})$  \\
			\hline
			\multicolumn{1}{c}{\multirow{5}{*}{N}}\\
			& Calc. & 909.75 & 5.57(4)& & 919.56 & 4.39(9)\\
			& Ref\cite{Kadrekar2011} & 909.475 & 9.158(4) &  & 919.113 & 2.409(9)\\
			& Ref\cite{Lyashchenko2021}& 910.03 & 1.44(5)&  & 919.65  & 4.24(9) \\
			\multicolumn{1}{c}{\multirow{6}{*}{Ne}}\\
			& Calc. & 1912.25 & 1.66（6） & &1926.95 & 9.49(9） \\
			& Ref\cite{Kadrekar2011} & 1911.507 & 1.343(6)&  & 1926.027 & 4.813(9) \\
			& Ref\cite{Lyashchenko2021} & 1912.14 & 2.28(6)&  & 1926.62 & 9.15(9) \\
			& Ref\cite{Goryayev2006} & & & &  1928.844 & 1.269(10)\\
			\multicolumn{1}{c}{\multirow{6}{*}{Fe}}\\
			& Calc.  & 13554.24 & 2.94(9) & & 13605.48 & 6.51(10) \\
			& Ref\cite{Kadrekar2011}  & 13553.011 & 1.599(9)  & & 13604.082 & 3.342(10) \\
			& Ref\cite{Lyashchenko2021} & 13554.14 & 3.08(9) & & 13065.13 & 6.88(10)\\
			& Ref\cite{Goryayev2006} & & & & 13625.14 & 9.42(10)\\
			\multicolumn{1}{c}{\multirow{4}{*}{Cu}}\\
			& Calc. & 16940.26 & 6.05(9)&  & 17004.20 & 8.13(10) \\
			& Ref\cite{Lyashchenko2021} & 16940.52 & 6.26(9)  & & 17004.24 & 8.54(10) \\
			\multicolumn{1}{c}{\multirow{4}{*}{Zn}}\\
			& Calc. & 18155.36 & 7.49(9) & & 18224.78 & 8.81(10) \\
			& Ref\cite{Lyashchenko2021} & 18155.27 & 7.75(9)  & & 18224.11 & 9.08(10) \\
			\multicolumn{1}{c}{\multirow{4}{*}{Ga}}\\
			& Calc. & 19413.86 & 9.22(9) & &  19488.49 & 9.43(10) \\
			& Ref\cite{Lyashchenko2021} & 19413.43 & 9.42(9)  & & 19487.83 & 9.68(10)\\
			\multicolumn{1}{c}{\multirow{4}{*}{Ge}}\\
			& Calc.  & 20715.94 & 1.11(10) & & 20796.59 & 9.94(10) \\
			& Ref\cite{Lyashchenko2021}   & 20715.19 & 1.13(10) & & 20795.69 & 1.03(11) \\
			\multicolumn{1}{c}{\multirow{4}{*}{As}}\\
			& Calc.  & 22061.97 & 1.32(10) & & 22149.31 & 1.06(11) \\
			& Ref\cite{Lyashchenko2021}  & 22062.53 & 1.33(10) & & 22149.70 & 1.09(11) \\
			\multicolumn{1}{c}{\multirow{4}{*}{Se}}\\
			& Calc.  & 23452.18 & 1.56(10) & & 23547.15 & 1.15(11) \\
			& Ref\cite{Lyashchenko2021} & 23452.32 & 1.56(10)&  & 23546.81 & 1.15(11)\\
			\multicolumn{1}{c}{\multirow{4}{*}{Br}}\\
			& Calc.  & 24886.51 & 1.84(10) & & 24989.13 & 1.19(11) \\
			& Ref\cite{Lyashchenko2021}  & 24886.35 & 1.82(10) & & 24988.82 & 1.22(11) \\
			\multicolumn{1}{c}{\multirow{4}{*}{Kr}}\\
			& Calc.  & 26365.42 & 2.13(10) & & 26476.75 & 1.26(11) \\
			& Ref\cite{Lyashchenko2021}  & 26364.84 & 2.11(10) & & 26476.03 & 1.29(11) \\
			\multicolumn{1}{c}{\multirow{4}{*}{Xe}}\\
			& Calc.  & 60925.15 & 1.44(11) & & 61385.51 & 3.19(11) \\
			& Ref\cite{Lyashchenko2021}  & 60925.75 & 1.17(11)& & 61385.80 & 2.92(11) \\
			\multicolumn{1}{c}{\multirow{4}{*}{W}}\\
			& Calc.  & 119021.40 & 3.83(11)& & 120721.90 & 4.77(11) \\
			& Ref\cite{Lyashchenko2021}  & 119023.02 & 3.56(11)& & 120723.21 & 6.40(11) \\
			\hline
		\end{tabular}\\
\end{table*}}

In Table \ref{Tab4}, we present the TEOP transition energy and probability of $2s2p-1s^2$, along with comparisons with other authors \cite{Kadrekar2011, Goryayev2006, Lyashchenko2021}. The transition energy of TEOP aligns well with the comparison data. The ratios of transition probabilities in Babushkin and Coulomb gauges are 0.96-1.12 and 0.87-1.32 for the OEOP and TEOP transition, respectively. Therefore, the atomic state wave function of the OEOP is demonstrated to be better than that of the TEOP by comparing the ratios under the two gauges.

\begin{figure}
\centering
\includegraphics[width=8.6cm]{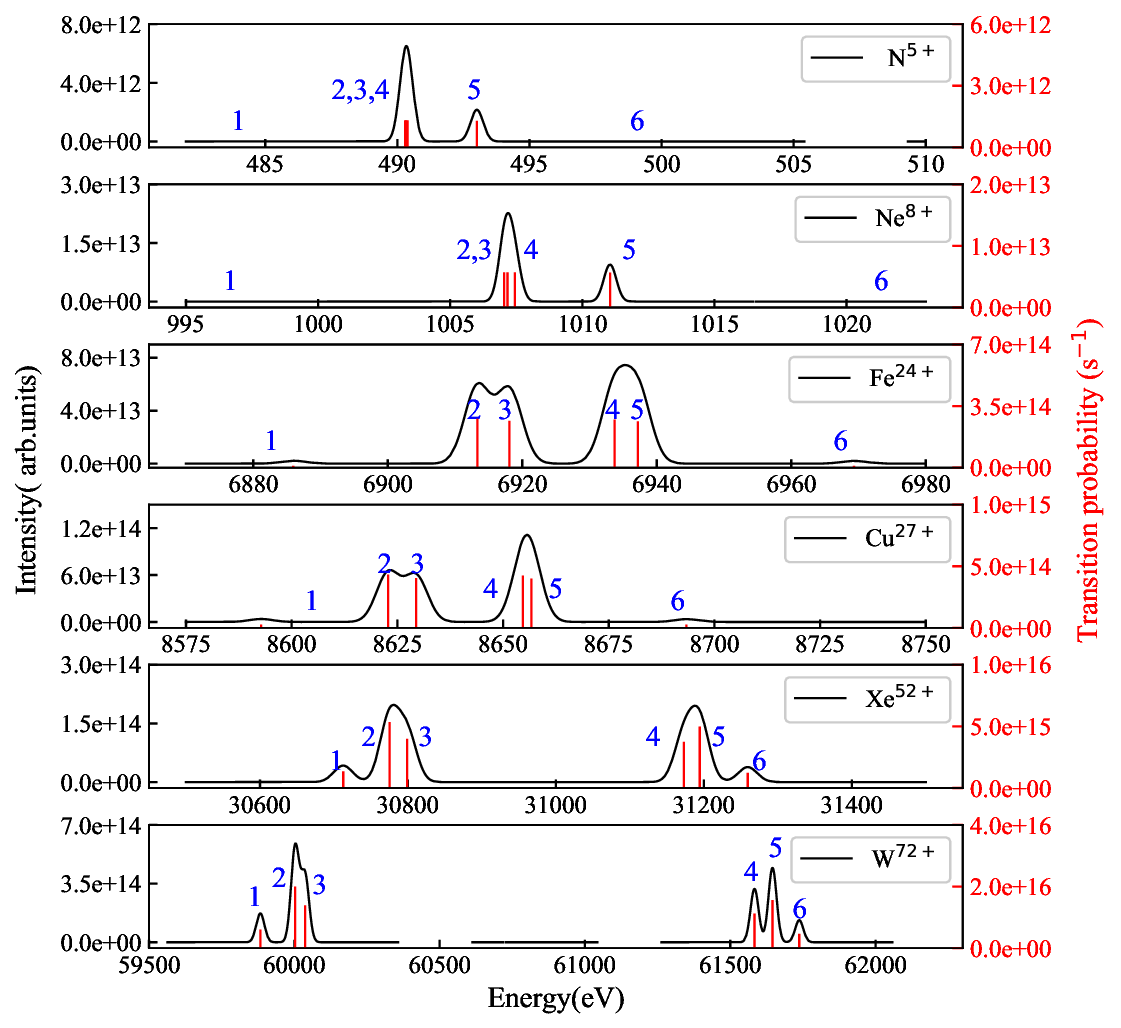}
\caption{The theoretical spectra of the OEOP transition for He-like ions from $2s2p-1s2s$. The blue curves is obtained by calculating the transition probability at different full width at half maximum (FWHM) according the experimental resolution. The red bar is the transition probability. Peak 1 - 6 denote $^3P_1$ - $^1S_0$, $^3P_0$ - $^3S_1$, $^3P_1$ - $^3S_1$, $^3P_2$ - $^3S_1$, $^1P_1$ - $^1S_0$, and  $^1P_1$ - $^3S_1$ transition, respectively. \label{fig1}}
\end{figure}

Based on the above calculation, the spectra of He-like ions OEOP and TEOP transition are predicted, which may provide helpful information for further experimental research. In Fig. \ref{fig1}, we present the theoretical spectra of the $2s2p-1s2s$ OEOP transition for several He-like ions. The spectral intensity is assumed to be proportional to the transition probability, and each individual transition has a Gaussian profile incorporating Doppler, natural, collision, and instrumental broadening effects. The Gaussian profile's full width at half maximum (FWHM) is adopted from the previous experimental resolution in different energy ranges. 

The high-energy X-ray transition has a Gaussian profile with an FWHM of 0.56 eV around 400-1000 eV, corresponding to the experimental resolution \cite{Nicolosi1977}. It is noteworthy that at a resolution of 0.56 eV, only two peaks could be observed for N$^{5+}$ and Ne$^{8+}$. However, these two peaks are not isolated transitions instead of blended transition liens. For instance, the left peak contains transitions $^3P_0$ - $^3S_1$, $^3P_1$ - $^3S_1$, $^3P_1$ - $^3S_1$, and the right peak is $^1P_1$-$^1S_0$ for N$^{5+}$. There are two spin-flip transitions $^3P_1$ - $^1S_0$ and $^1P_1$ - $^3S_1$ for N$^{5+}$ with transition energies 483.20 and 500.14 eV, respectively, which do not appear due to their small intensity.

In the 6 keV - 8 keV energy region, the FWHM is 4.5 eV and 6.5 eV for Fe and Cu, respectively \cite{Porter2009, Porter2004}. For Fe and Cu, the $^3P_0$ - $^3S_1$, $^3P_1$ - $^3S_1$ transitions are in the middle peak, and the $^3P_2$ - $^3S_1$ and $^1P_1$ - $^1S_0$ are in the right peak. Two forbidden transitions still show weaker intensities.

With an energy resolution of 30 eV at 30 keV \cite{Porter2009}, Xe exhibits four peaks. From left to right, the first and fourth peaks are the forbidden transitions of $^3P_1$ - $^1S_0$ and $^1P_1$ - $^3S_1$, respectively. The second peak contains $^3P_0$ - $^3S_1$, $^3P_1$ - $^3S_1$, and the third peak contains $^1P_1$ - $^1S_0$ and $^3P_2$ - $^3S_1$. Inversion of the transition energy of $^3P_2$ - $^3S_1$ and $^1P_1$ - $^1S_0$ is observed in the transition line enveloped by the third peak around 31.2 keV.

For W, five peaks are observed, making the energy region better resolved at an FWHM of 33 eV around 60 keV \cite{Porter2009}. In this case, the inversion between the $^3P_2$ - $^3S_1$ and $^1P_1$ - $^1S_0$ transitions become more apparent.

The effect of Breit interaction on multiple state splitting in most double-hole states is significant \cite{Chen1982}. Further analysis reveals that the Breit interaction also plays a vital role in the inversion of the $^1P_1$ - $^1S_0$ and $^3P_2$ - $^3S_1$ transition energies. In Table \ref{Tab5}, we calculate the Breit, QED, and Coulomb contribution to the transition energy of each ion. 

The analysis indicates that the structure of middle Z ions (Z = 33-36) changes under the influence of Breit interaction. At the same time, the transition energy remains unchanged under Coulomb interaction and QED effects. This is because the Breit interaction reduces the binding energy for each state of the $2s2p$ configuration and the $^1S_0$ state of the $1s2s$ configuration but slightly increases the binding energy for the $^3S_1$ state of the $1s2s$ configuration.

\begin{figure}
\centering
\includegraphics[width=8.6cm]{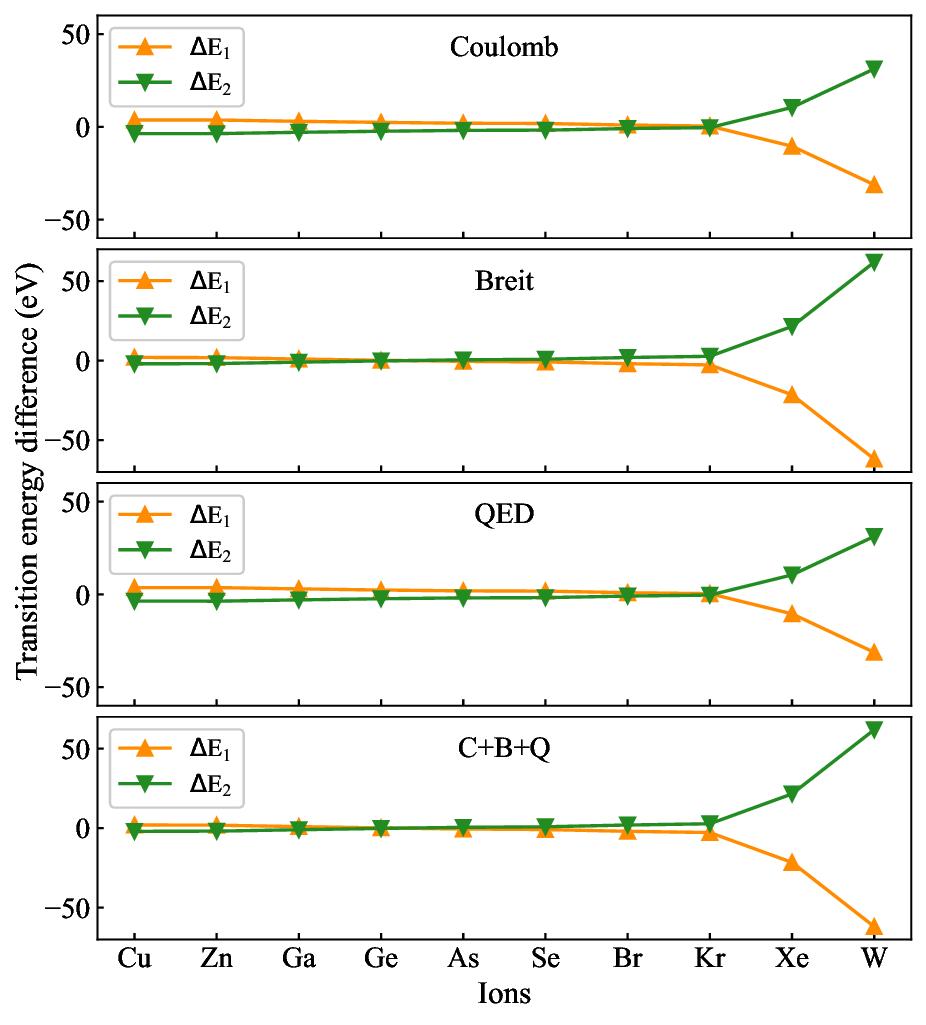}
\caption{The Coulomb, Breit, QED and total contribution to the transition energy of $^1P_1$ - $^1S_0$ and $^3P_2$ - $^3S_1$ transition of $2s2p-1s2s$ of He-like ions. $\Delta$E$_1$ is the transition energy difference of $^1P_1$ - $^1S_0$ and $^3P_2$ - $^3S_1$, $\Delta$E$_1$=-$\Delta$E$_2$.\label{fig2}}
\end{figure}

However, at high Z (Z=54 and 74) ions, the transition energy of $^1P_1$-$^1S_0$ is greater than that of $^3P_2$-$^3S_1$ under all three corrections. This is attributed to the significant effects of Coulomb interaction, QED effect, and Breit interaction at high Z. To demonstrate this variation, Figure \ref{fig2} is shown.

\begin{table*}
	\small
	\caption{The Coulomb, Breit, QED contribution to the transition energy (in eV) of OEOP transition $2s2p-1s2s$ of He-like ions. C+B+Q include the sum of Coulomb, Breit and QED effects, while QED and Breit include the Coulomb contribution simultaneously. $\Delta$QED, $\Delta$Breit, and $\Delta$(B+Q) are the QED, Breit, and total contribution to the transition energy, respectively. \label{Tab5}}
	\centering
	\begin{tabular}{cccccccc}
		\hline
		\noalign{\smallskip}
		& & & \multicolumn{2}{c}{$^3P_2$ - $^3S_1$} & & &  \\
		\hline
		&   C+B+Q & Coulomb & QED  &  Breit & $\Delta$(B+Q) &$\Delta$(QED) &$\Delta$(Breit) \\
		\hline
		Cu &8654.69 & 8660.17 & 8654.63 & 8660.22 & -5.48 & -5.54 & 0.05\\
		Zn &9272.57	& 9278.70 & 9272.51 & 9278.76	& -6.13	& -6.19	& 0.06\\
		Ga &9912.63	& 9919.47 & 9912.57	& 9919.54 & -6.84 & -6.9 & 0.07\\
		Ge & 10575.28 & 10582.76 & 10575.21	&10582.83 & -7.48 & -7.55 & 0.07\\
		As &11260.43 & 11268.71 & 11260.35	& 11268.79 & -8.28& -8.36 & 0.081\\
		Se &11968.38 & 11977.51 & 11968.29 & 11977.60 & -9.13 & -9.22 & 0.09\\
		Br &12699.09 & 12709.14	& 12699.00	& 12709.23 & -10.05 & -10.14 & 0.09\\
		Kr &13452.91 & 13463.94	& 13452.81 & 13464.04 & -11.03	& -11.13 & 0.1\\
		Xe &31194.45 & 31235.69	& 31194.13 & 31236.01 & -41.24	& -41.56 & 0.32\\
		W & 61645.39 & 61760.55	& 61644.68 & 61761.26 & -115.16 & -115.87 & 0.71\\
		\hline
		\noalign{\smallskip}
		& & & \multicolumn{2}{c}{$^1P_1$ - $^1S_0$} & & &  \\
		\hline
		&   C+B+Q & Coulomb & QED  &  Breit & $\Delta$(B+Q) &$\Delta$(QED) &$\Delta$(Breit) \\
		\hline
		Cu &8656.7 & 8663.81 & 8658.27 & 8662.24 & -7.11 & -5.54 & -1.57\\
		Zn & 9274.44 & 9282.37	& 9276.18 & 9280.64	& -7.93	& -6.19	& -1.73\\
		Ga & 9913.61 & 9922.43	& 9915.53 & 9920.51 & -8.82	& -6.9	& -1.92\\
		Ge & 10575.44 & 10585.15 & 10577.56	& 10583.03	& -9.71 & -7.59	& -2.12\\
		As & 11259.91 & 11270.63 & 11262.24	& 11268.31	& -10.72 & -8.39 & -2.32\\
		Se & 11967.51 & 11979.30 & 11970.05	& 11976.76	& -11.79 & -9.25 & -2.54\\
		Br & 12697.11 & 12710.08 & 12699.90	& 12707.29	& -12.97 & -10.18 & -2.79\\
		Kr & 13450.15 & 13464.36 & 13453.19	& 13461.33	& -14.21 & -11.17 & -3.031\\
		Xe & 31172.90 & 31225.23 & 31183.63 & 31214.51	& -52.33 & -41.60 & -10.72\\
		W & 61583.47 & 61729.33	& 61613.44	& 61699.39	& -145.86 & -115.89 & -29.94\\
		\hline
	\end{tabular}
\end{table*}
%

In Fig. \ref{fig2}, the difference of the transition energies of $^1P_1$-$^1S_0$ and $^3P_2$-$^3S_1$ with each correction from Table \ref{Tab5} are shown. $\Delta E_1$ is defined as the $^3P_2$-$^3S_1$ transition energy minus that of $^1P_1$-$^1S_0$, while $\Delta E_2$ = -$\Delta E_1$. As the atomic number increases, the transition energy of $^1P_1$-$^1S_0$ decreases with the inclusion of the Breit interaction. In contrast, the transition energy of $^3P_2$-$^3S_1$ increases with the Breit interaction, and the transition energy inversion occurs at the As ion. 

In the case of Xe and W, the $^1P_1$-$^1S_0$ transition energy is always less than $^3P_2$-$^3S_1$ under the influence of all three effects.

\begin{figure}
\centering
\includegraphics[width=8.6cm]{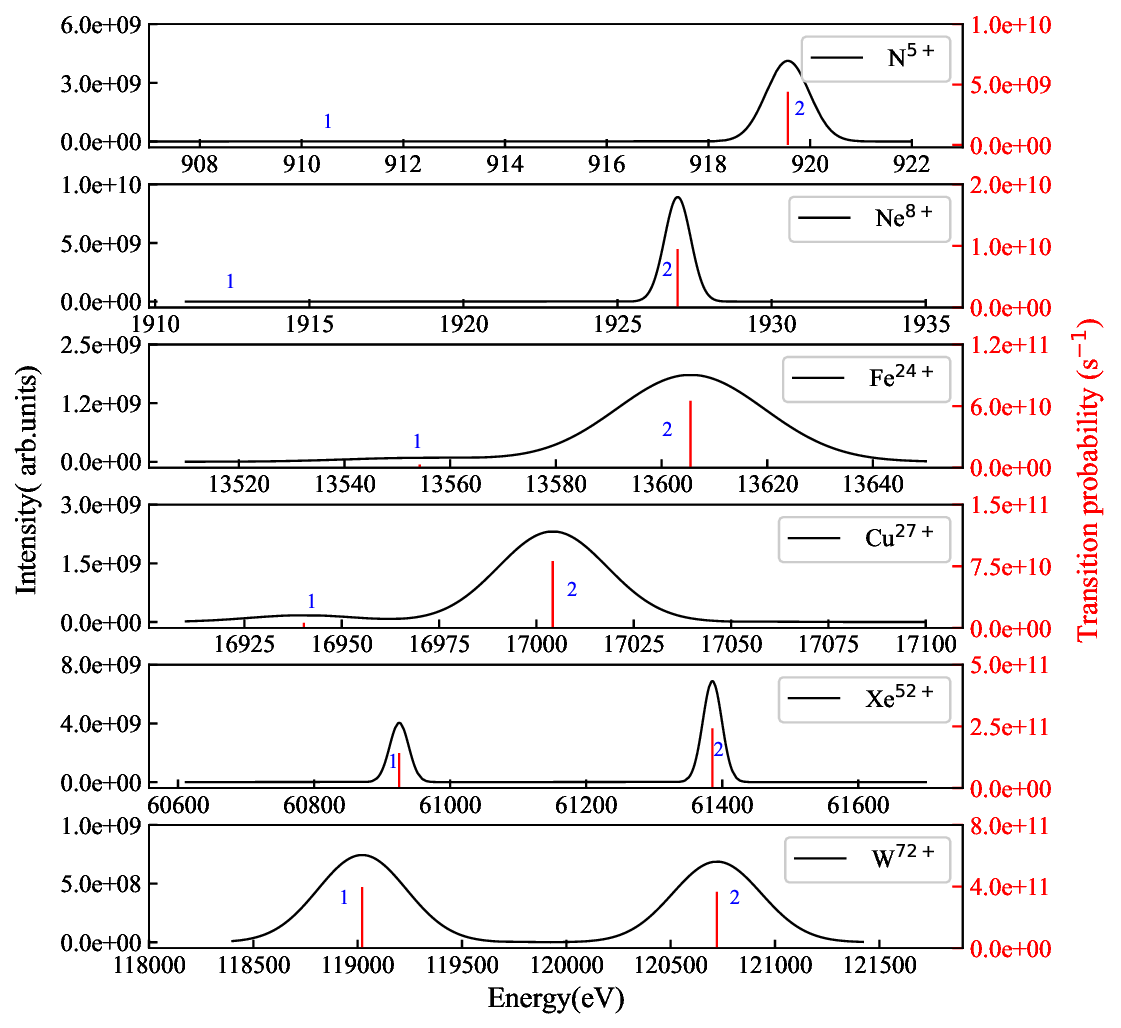}
\caption{The theoretical spectra of the TEOP transition for He-like ions from $2s2p-1s^2$. The blue curves is obtained by calculating the transition probability at different full width at half maximum (FWHM) according to the experimental resolution. The red bar is the transition probability. 1 and 2 denote $^3P_1$ - $^1S_0$, $^1P_1$ - $^1S_0$ transitions, respectively.\label{fig3}}
\end{figure}

In FIG. \ref{fig3}, we present the spectra intensity of the TEOP transition for He-like ions from $2s2p-1s^2$ transition. TEOP transition spectra were obtained at different FWHM in different X-ray energy ranges. The spectra of N$^{5+}$ and Ne$^{8+}$ have a resolution of 1 eV \cite{Andriamonje1991, Lamy1990}, with no peak due to the small transition probability of a $^3P_1$ - $^1S_0$ transition and low intensity.

The FWHM of the spectra of Fe$^{24+}$ and Cu$^{27+}$ ions, whose energies range from 13-16 keV, were set at 33 eV. After broadening the transition probability, it is found that the transition probability of the forbidden transition is still tiny, resulting in an insignificant peak. However, the allowed transition line could be observed.

For Xe$^{52+}$, the energy resolution is 33 eV at 60 keV \cite{Porter2009}, and two peaks appear. Generally, the full width at half maximum for energy of 121 keV is 500 eV \cite{Shao2017}. The resulting spectrum of the W$^{72+}$ ion with an FWHM of 500 eV is given. 

In Fig \ref{fig3}, $^3P_1$-$^1S_0$ is spin forbidden. Generally, its intensity is low, causing the peak not to appear. However, as the atomic number increases, its intensity increases. Especially for the case in W$^{72+}$ ion, the forbidden transition has comparable intensity with the allowed transition. 

\section{CONCLUSIONS}

In this study, we employed a comprehensive approach to calculate the transition energies and probabilities for the one-electron one-photon (OEOP) and two-electron one-photon (TEOP) processes in He-like ions, specifically for N, Ne, Fe, Cu, Ga, Ge, As, Se, Br, Kr, Xe, and W ions. Our calculations were performed using the fully relativistic multi-configuration Dirac-Hartree-Fock (MCDHF) method in conjunction with the active space approach, incorporating the GRASP2K package. This code allowed us to systematically consider relativistic effects, electron correlation, Breit interaction, and quantum electrodynamics effects.

To accurately consider the electron correlation effects, we developed suitable electron correlation models, ensuring our calculations align well with existing results. Furthermore, according to the existing experiments, we treated each transition as a Gaussian profile, accounting for Doppler, natural, collisional, and instrumental broadening effects. This approach enabled us to predict the OEOP and TEOP transition spectra for He-like ions at varying resolutions.

Our results exhibited excellent agreement with previous works, confirming the reliability of our methodology. Notably, in the OEOP transition, we observed an inversion of the $^1P_1$ - $^1S_0$ and $^3P_2$ - $^3S_1$ transitions at As influenced by the increasing atomic number and the impact of Breit interaction. Additionally, the spin-forbidden transitions, $^3P_1$ - $^1S_0$ and $^1P_1$ - $^3S_1$, displayed a progressive increase in intensity with atomic number.

In the TEOP transition, the $^3P_1$ - $^1S_0$ transition emerged as forbidden, while $^1P_1$ - $^1S_0$ stood out as the principal decay channel. This insight enhances our theoretical understanding of these exotic processes and serves as a valuable reference for experimental investigations. Our comprehensive approach sheds light on the intricate interplay of relativistic effects, electron correlation, and quantum electrodynamics phenomena in He-like ions, contributing to the broader knowledge of atomic physics.




\begin{acknowledgments}
This work was supported by the National Natural Science Foundation
of China (Grant No. 12274352), National Key Research, and
Development Program of China (Grant No. 2022YFA1602500).
\end{acknowledgments}

\bibliography{zhao.bib}

\end{document}